\documentclass[a4paper]{jpconf}
\usepackage{graphicx,bm}
\usepackage{amsmath,amssymb}
\usepackage{cite}
\usepackage[dvips]{color}

\begin{document}
\title{Ground-state properties of a triangular triple quantum dot 
connected to superconducting leads
}

\author{Akira Oguri,$^1$
Izumi Sato,$^1$ Masashi Shimamoto,$^1$ and Yoichi Tanaka$^2$}

\address{$^{1}$ Department of Physics, Osaka City University, 
Osaka 558-8585, Japan}

\address{$^{2}$ Condensed Matter Theory Laboratory, RIKEN, 
Saitama 351-0198, Japan}


\begin{abstract}
We study ground-state properties of a triangular 
triple quantum dot connected to two superconducting (SC) leads. 
In this system orbital motion along the triangular configuration 
causes various types of quantum phases, 
such as a Kondo effect with a four-fold degenerate state 
and the Nagaoka ferromagnetic mechanism, 
depending on the electron filling.  
The ground state also evolves 
as the Cooper pairs penetrate from the SC leads.
We describe the phase diagram 
in a wide range of the parameter space, 
varying the gate voltage, 
the couplings between the dots and leads, 
and also the Josephson phase between the SC gaps. 
The results are obtained in the limit of large SC gap,   
carrying out exact diagonalization of an effective Hamiltonian.
We also discuss 
a classification of the quantum states   
 according to the fixed point of the Wilson numerical renormalization group 
(NRG). 
Furthermore, we show that the Bogoliubov zero-energy excitation  
determines the ground state of a $\pi$ Josephson junction 
at small electron fillings.

\end{abstract}

\section{Introduction}

\begin{figure}[b]

\ \hspace{-1.3cm}
\begin{minipage}{0.45\linewidth}
\setlength{\unitlength}{0.75mm}

\begin{picture}(110,40)(0,0)
\thicklines

\put(17,1){\line(1,0){22}}
\put(17,15){\line(1,0){22}}
\put(39,1){\line(0,1){14}}

\put(17.5,4){\makebox(0,0)[bl]{\large SC:$\,\Delta_L^{\phantom{\dagger}}$}}

\put(76,1){\line(1,0){22}}
\put(76,15){\line(1,0){22}}
\put(76,1){\line(0,1){14}}

\put(78,4){\makebox(0,0)[bl]{\large SC:$\,\Delta_R^{\phantom{\dagger}}$}}

\put(51,28){\line(0,1){12}}
\put(65,28){\line(0,1){12}}
\put(51,28){\line(1,0){14}} 

\put(55,31.5){\makebox(0,0)[bl]{\large $N$}}

\multiput(58,27.2)(0,-2){5}{\line(0,-1){1}}
\multiput(40.0,7.5)(2,0){5}{\line(1,0){1}}
\multiput(66,7.5)(2,0){5}{\line(1,0){1}}

\put(58,17.5){\circle*{4.5}} 
\put(52,7.5){\circle*{4.5}} 
\put(64,7.5){\circle*{4.5}} 


\put(54,7.5){\line(1,0){8}}
\put(51.5,7.5){\line(3,5){7}}
\put(64.5,7.5){\line(-3,5){7}}


\put(60,18){\makebox(0,0)[bl]{\large $\Gamma_N^{}$}}
\put(42,8.75){\makebox(0,0)[bl]{\large $\Gamma_L^{}$}}
\put(66,8.75){\makebox(0,0)[bl]{\large $\Gamma_R^{}$}}
\end{picture}
\end{minipage}
\rule{0.03\linewidth}{0cm}
\raisebox{0.05cm}{
\begin{minipage}{0.54\linewidth}
\caption{\baselineskip=0.77\baselineskip
Triangular triple quantum dots ({\large $\bullet$}) 
coupled to one normal ($N$) and two superconducting leads ($L$,$R$)
with the gaps $\Delta_{L/R} =|\Delta_{L/R}|\, e^{i\theta_{L/R}}$.
Here, $\Gamma_{\nu} \equiv \pi \rho \,v^{2}_{\nu}$ 
with $\rho$ the density of states 
of the leads $\nu=L,R,N$, and 
$v_{\nu}$ the tunneling matrix element.  
}
\end{minipage}
}%
\label{fig:system}
\end{figure}
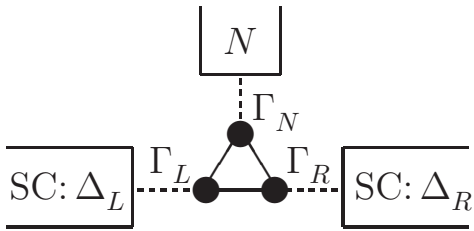


Triangular triple quantum dot (TTQD) is one of the 
multi-dot systems that has intensively been studied in these years. 
One of the interesting features of this system is 
rich internal degrees of freedom 
that emerge through orbital motion along the triangular 
configuration and also from several different possible concentrations 
of occupying electrons. Furthermore, 
the number of the conducting channels 
and the way the leads are connected to the TTQD  
also give a variety to ground state and low-energy excitations. 
For these reasons, various kinds of Kondo effects and quantum phase 
transition have been predicted for this system 
\cite{KKA1,ONTN,Zitko2,Logan3,Ulloa,Numata2,aoTTQD2011,Koga}.

This report focuses on the interplay and competition 
between the electron correlation and superconductivity 
\cite{MartinYeyati,sc3,TanakaNQDS,Deacon,MaurandSchonenberger}. 
Specifically,  we consider the TTQD that is embedded 
in a Josephson junction between two 
superconducting (SC) leads  as illustrated 
in Fig.\ \ref{fig:system} \cite{GovePalaKonig_PRB,aoYoichiBauer}. 
We explore the wide parameter space of this junction  
to clarify how the Copper pairs penetrating into 
the TTQD realize various quantum phases in the presence of 
a large Coulomb interaction.

\section{Model \& Formulation}

We start with the Hamiltonian, given in the form   
$\, H =  H_\mathrm{QD}^{} + H_\mathrm{T}^{}  + H_\mathrm{lead}\,$ with 
\begin{align}
& 
H_\mathrm{QD}^{} \,= \,
-\,t\, 
\sum_{\langle ij \rangle}
\sum_\sigma 
\left(
d_{i\sigma}^{\dag}d_{j\sigma}^{}
+ d_{j\sigma}^{\dag}d_{i\sigma}^{}
\right)
\,+\, \sum_{i}
\sum_{\sigma} \epsilon_{d}\, n_{d,i\sigma}
\,+\,U \sum_{i}
n_{d,i\uparrow}^{}n_{d,i\downarrow}^{} \;, 
\\
& 
H_\mathrm{T}^{} \,= \,
\sum_{\nu =N,L,R} 
 \sum_{\sigma}  v_{\nu}^{}
 \left( \psi_{\nu\sigma}^\dag d_{\nu\sigma}^{} + 
  d_{\nu\sigma}^{\dag} \psi_{\nu\sigma}^{} \right) , 
\qquad \quad 
\psi_{\nu\sigma}^{} \equiv 
\int_{-D}^D \! d\varepsilon \, \sqrt{\rho} 
\, c^{}_{\varepsilon\alpha m} 
\\
& H_\mathrm{lead} \,=\,   \sum_{\nu=N,L,R}
\sum_\sigma
\int_{-D}^D  \! d\varepsilon\,  \varepsilon\, 
 c^{\dagger}_{\varepsilon \nu \sigma} c_{\varepsilon \nu \sigma}^{} 
%
\ + 
\sum_{\alpha = L,R} 
\int_{-D}^D  \! d\varepsilon\,  
\left(\Delta_{\alpha}\, 
c_{\varepsilon\alpha\uparrow}^\dag \,c_{\varepsilon\alpha\downarrow}^\dag 
+ \textrm{H.c.}\right) . 
\label{eq:H}
\end{align}
%
where $d^{\dag}_{i\sigma}$  is the creation operator 
an electron with energy $\epsilon _{d}$ and spin $\sigma$ 
in the quantum dot at site $i$, $U$ is  
the Coulomb interaction $U$ and
$n_{d,i\sigma} \equiv d^{\dag}_{i\sigma}d^{}_{i\sigma}$. 
Similarly, $c_{\varepsilon\nu\sigma}^\dag$ creates an electron  
in the lead $\nu =L,\,R,\,N$, and is normalized as
$\{ c^{}_{\varepsilon\nu \sigma}, 
c^{\dagger}_{\varepsilon'\nu'\sigma'}
\} = \delta_{\nu\nu'} \delta_{\sigma\sigma'}   
\delta(\varepsilon-\varepsilon')$. 
We assign the same label  for the dot,   
 $i=L,\,R,\,N$, as the one for the adjacent lead that is coupled 
via the tunneling matrix element $v_{\nu}^{}$ 
with $\rho=1/(2D)$ and  $\Gamma_{\nu} \equiv \pi \rho\, v_{\nu}^{2}$.  
The superconducting gap  
 $\Delta_{\alpha}^{} = |\Delta_{\alpha}^{}| e^{i\theta_{\alpha}}$ 
induces the Josephson current between the two SC leads 
for finite phase differences $\phi \equiv \theta_{R}^{} -\theta_{L}^{}$.

 Since this Hamiltonian has a number parameters to be explored,
we consider the case where  
 $\Gamma_L^{} = \Gamma_R^{}$ $(\equiv \Gamma_S)$ and  
$|\Delta_L^{}| = |\Delta_R^{}|$ $(\equiv \Delta_\mathrm{SC})$.  
Specifically, we concentrate on the case   
where the SC gap is much larger than the other energy scales
 $\Delta_\mathrm{SC} \gg 
\max(\Gamma_{S}, \Gamma_N, U, |\epsilon_d|)$  but $D$, 
and the quasi-particle excitations 
in the continuum energy region above the SC gap are projected out. 
Nevertheless, the essential physics of the SC proximity is still preserved, 
and in the limit of $\Delta_\mathrm{SC} \to \infty$  
it is described by an effective Hamiltonian 
\begin{align}
& H_\mathrm{eff}^{} \,=\, 
 H_\mathrm{QD}^{} +\, 
\sum_{\alpha = L,R} 
\left(
\Delta_{d,\alpha}^{}\, 
d_{\alpha\uparrow}^\dag \,d_{\alpha\downarrow}^\dag 
+ 
\Delta_{d,\alpha}^*\, 
d_{\alpha\downarrow}^{} \,d_{\alpha\uparrow}^{} 
\right) ,   
\qquad \quad 
\Delta_{d,\alpha} \equiv  
\  \Gamma_S \,e^{i\theta_{\alpha}^{}} . 
\label{eq:H_eff}
\end{align}
The static SC pair potential  $\Delta_{d,\alpha}$ 
is induced in the TTQD, 
the amplitude of which is determined by the coupling strength 
$|\Delta_{d,\alpha}| \equiv \Gamma_S$  
while the phase $\theta_\alpha$ preserves 
that of the SC hosts \cite{Cuevas,sc2}.
In the present report, we concentrate on the $\Gamma_N=0$ case 
where the normal lead is disconnected.

\begin{figure}[t] 
 \leavevmode

\rule{-0.02\linewidth}{0cm}
\begin{minipage}{1\linewidth}
\includegraphics[width=0.245\linewidth]{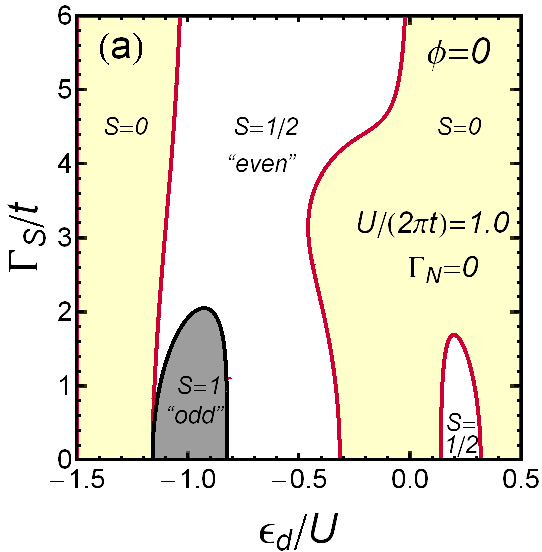}
\includegraphics[width=0.245\linewidth]{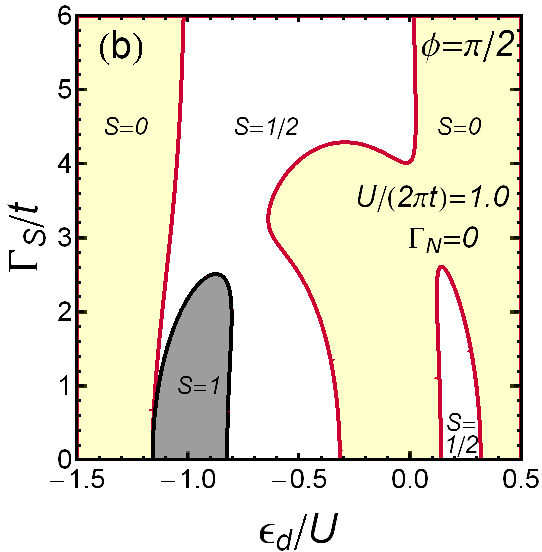}
\includegraphics[width=0.245\linewidth]{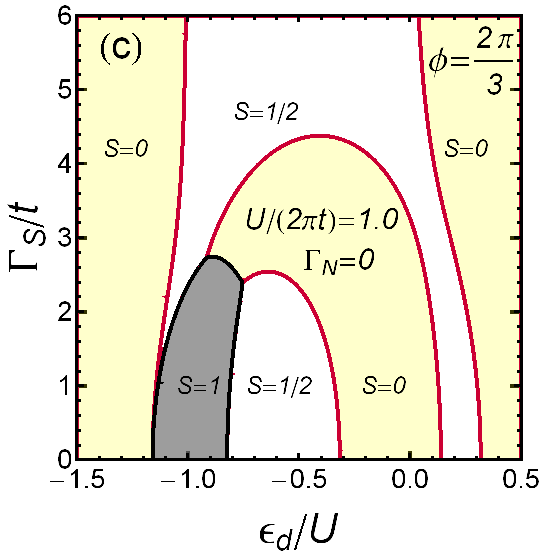}
\includegraphics[width=0.245\linewidth]{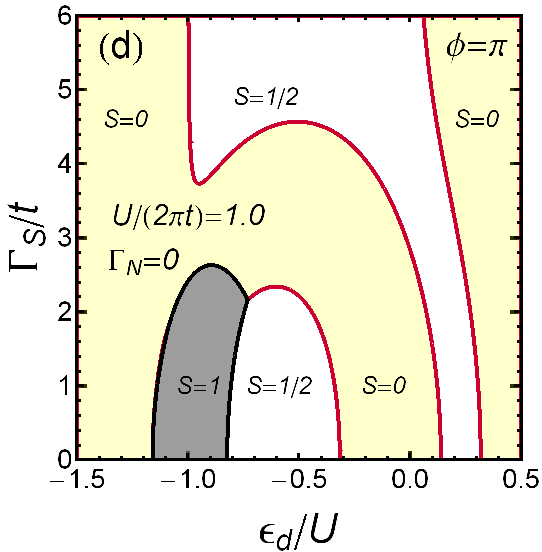}
\end{minipage}
\caption{
%
Ground-state phase diagram of the TTQD for $\Gamma_N=0$, is plotted 
in an $\epsilon_d$ vs $\Gamma_S$ 
plane for 
 (a) $\phi = 0$,  (b) $\phi = \pi/2$, (c) $\phi = 2\pi/3$,   
and (d) $\phi = \pi$,   
 for relatively large Coulomb interaction $U=2\pi t$.  
The eigenstates can be classified according 
to the total spin $S$, which takes values of  
$S=0$, $1/2$, and $1$ in the yellow, white, and gray regions, respectively.   
} 
\label{fig:results_2SC_terminal}
\end{figure}

\section{Ground states \& Fixed points for the TTQD connected to two SC leads}

\begin{figure}[t]
\rule{0.01\linewidth}{0cm}
\includegraphics[width=0.95\linewidth]{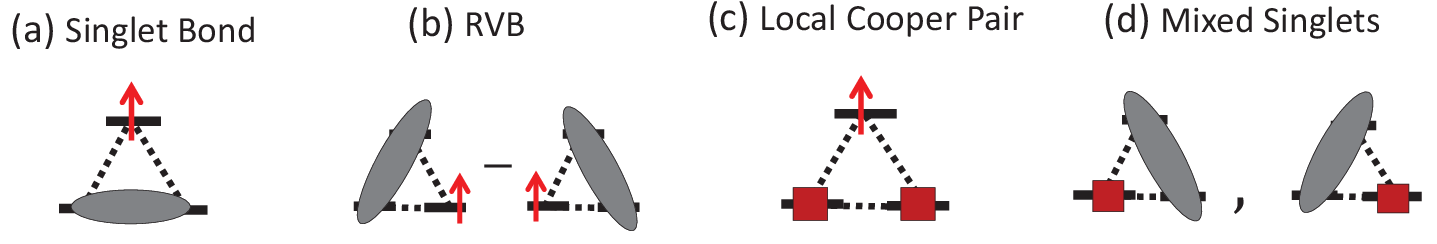}
%
\caption{
\baselineskip=0.5\baselineskip
Schematic pictures of the 
fixed-point wavefunctions near half-filling,  
 defined 
in Eqs.\ \eqref{eq:SB_fixed_point}--\eqref{eq:Mixed_singlet_fixed_point} 
 for $\Gamma_N^{} =0$.
The gray oval denotes the singlet bond, and the brown square denotes the 
local Cooper pair (LCP).
The mixed singlets (d) have two independent configurations. 
}
\label{fig:fixed_points}
\end{figure}

In this section, we describe 
 how the ground state of the TTQD coupled to two SC leads 
evolves as the Hamiltonian parameters $\epsilon_d$ and $\Gamma_S$ vary. 
Figure \ref{fig:results_2SC_terminal} shows the 
phase diagram for the ground state of $H_\mathrm{eff}^{}$ 
for relatively large interaction $U=2\pi t$.  
In the horizontal direction 
the average number of electrons $N_\mathrm{tot}$ 
in the TTQD increases as $\epsilon_d$ decreases,  
while in the vertical direction the SC proximity becomes large 
 as $\Gamma_S$ increases.
Specifically, for $\Gamma_S=0$, 
the occupation discontinuously changes   
near $\epsilon_d/U \simeq -1.2, -0.8,-0.3, 0.1$, and $0.3$. 
Correspondingly, level crossing occurs between  
the states of different total spins, 
 $S=1/2$ and $S=0$, 
or $S=1$ Nagaoka state that is caused by orbital motions 
along the triangular 
configuration \cite{ONTN,Numata2,aoTTQD2011}.
We see that the ground state evolves sensitively to 
the phase difference $\phi$ for intermediate values of 
the SC proximity, around $\Gamma_S^{}/t \simeq 3.0$, 
near half-filling $\epsilon/U \simeq -0.5$  
where $N_\mathrm{tot}\simeq 3.0$. 
This reflects properties of the wavefunctions, 
which are schematically illustrated in Fig.\ \ref{fig:fixed_points} 
and can be expressed in the form 
\begin{align}
\left| \Phi_{SB} \right\rangle
\ =& \  
d_{N\uparrow}^{\dag}
\left(
d_{L\uparrow}^{\dag} d_{R\downarrow}^{\dag}
-
d_{L\downarrow}^{\dag} d_{R\uparrow}^{\dag}
\right)
\left| 0 \right\rangle
\;,
\label{eq:SB_fixed_point}
\\
\left| \Phi_{RVB} \right\rangle
\,=& \  
\frac{1}{\sqrt{6}}
\left\{d_{R\uparrow}^{\dag}
\left(
d_{N\uparrow}^{\dag} d_{L\downarrow}^{\dag}
-
d_{N\downarrow}^{\dag} d_{L\uparrow}^{\dag}
\right)
-d_{L\uparrow}^{\dag}
\left(
d_{N\uparrow}^{\dag} d_{R\downarrow}^{\dag}
-
d_{N\downarrow}^{\dag} d_{R\uparrow}^{\dag}
\right)\right\}
\left| 0 \right\rangle 
\nonumber \\
= & \ 
\frac{1}{\sqrt{6}}
\left( 
  d_{N\uparrow}^{\dag} d_{L\downarrow}^{\dag} d_{R\uparrow}^{\dag}
+ d_{N\uparrow}^{\dag} d_{L\uparrow}^{\dag} d_{R\downarrow}^{\dag}
- 2d_{N\downarrow}^{\dag} d_{L\uparrow}^{\dag} d_{R\uparrow}^{\dag}
\right)
\left| 0 \right\rangle  \;, 
\label{eq:RVB_fixed_point}
\\
\left| \Phi_{LCP} \right\rangle
\,=& \  
d_{N\uparrow}^{\dag}
\left(
e^{-i\frac{\theta_L}{2}}
-e^{i\frac{\theta_L}{2}}
d_{L\uparrow}^{\dag}d_{L\downarrow}^{\dag}
\right)
\left(
e^{-i\frac{\theta_R}{2}}
-e^{i\frac{\theta_R}{2}}
d_{R\uparrow}^{\dag}d_{R\downarrow}^{\dag}
\right)
\left| 0 \right\rangle \;,
\label{eq:LCP_fixed_point}
\\
\left| \Phi_{\mathrm{mix},1} \right\rangle
\,=& \  
\frac{1}{2}
\left(
e^{-i\frac{\theta_L}{2}}
-e^{i\frac{\theta_L}{2}}
d_{L\uparrow}^{\dag}d_{L\downarrow}^{\dag}
\right)
\left(
d_{N\uparrow}^{\dag} d_{R\downarrow}^{\dag}
-
d_{N\downarrow}^{\dag} d_{R\uparrow}^{\dag}
\right)
\left| 0 \right\rangle \;,
\\
\left| \Phi_{\mathrm{mix},2} \right\rangle
\,=& \  
\frac{1}{2}
\left(
e^{-i\frac{\theta_R}{2}}
-e^{i\frac{\theta_R}{2}}
d_{R\uparrow}^{\dag}d_{R\downarrow}^{\dag}
\right)
\left(
d_{N\uparrow}^{\dag} d_{L\downarrow}^{\dag}
-
d_{N\downarrow}^{\dag} d_{L\uparrow}^{\dag}
\right)
\left| 0 \right\rangle \;. 
\label{eq:Mixed_singlet_fixed_point}
\end{align}
These states also represent 
typical fixed points of the Wilson numerical renormalization group 
(NRG) near half-filling. 
Specifically, the two spinful states,  
$\left| \Phi_{SB} \right\rangle$ and $\left| \Phi_{RVB} \right\rangle$, 
constitute 
a four-fold degenerate  ground state  
for $\Gamma_S=0$ in the strong interaction case $U\gg t$ 
\cite{KKA1,ONTN,Zitko2,Logan3,Ulloa,Numata2,aoTTQD2011,Koga}.
Note that the SB and RVB can be categorized into the even and odd  
parity states, respectively, 
with respect to a vertical axis 
passing through the apex site of the triangle (see Fig.\ \ref{fig:system}).

\begin{figure}[t] 
 \leavevmode
\rule{0.03\linewidth}{0cm}
\begin{minipage}{0.9\linewidth}
\includegraphics[width=0.48\linewidth]{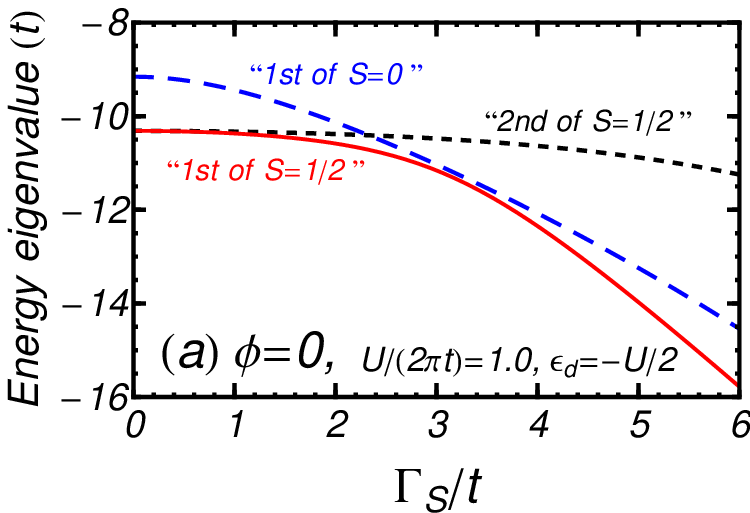}
\rule{0.02\linewidth}{0cm}
\includegraphics[width=0.48\linewidth]{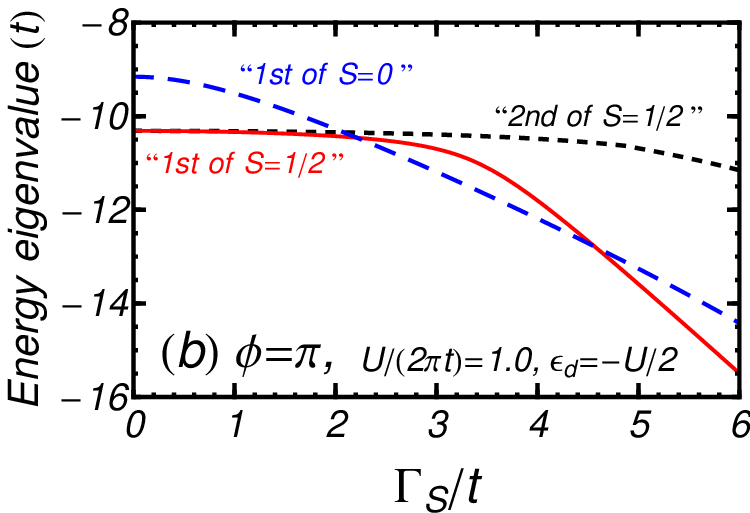}
\end{minipage}
\caption{
Eigenvalues of $H_\mathrm{eff}$ vs $\Gamma_S$, 
for  (a) $\phi = 0$ and  (b) $\phi = \pi$, 
at $\epsilon_d =-U/2$ and $U=2\pi t$. 
First two energies are shown for the subspace of spin $S=1/2$,  
while only the lowest one is shown for $S=0$. 
} 
 \label{fig:energy}
\end{figure}

\begin{figure}[t] 
 \leavevmode

\rule{0.03\linewidth}{0cm}
\begin{minipage}{0.9\linewidth}
\includegraphics[width=0.48\linewidth]{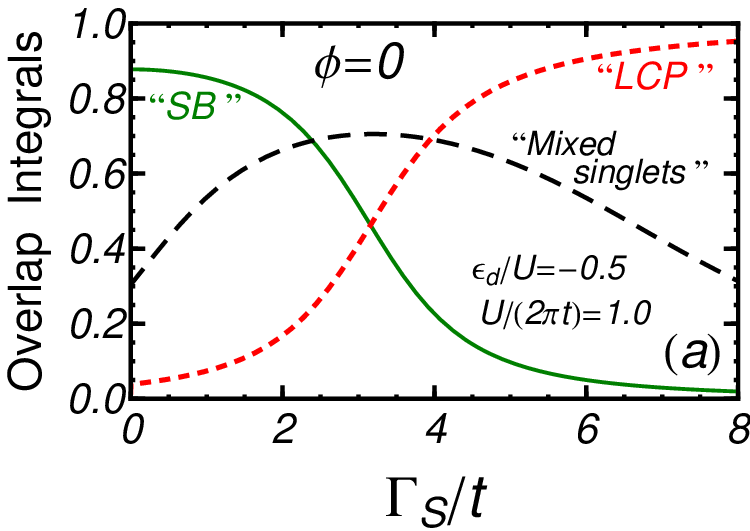}
\rule{0.02\linewidth}{0cm}
\includegraphics[width=0.48\linewidth]{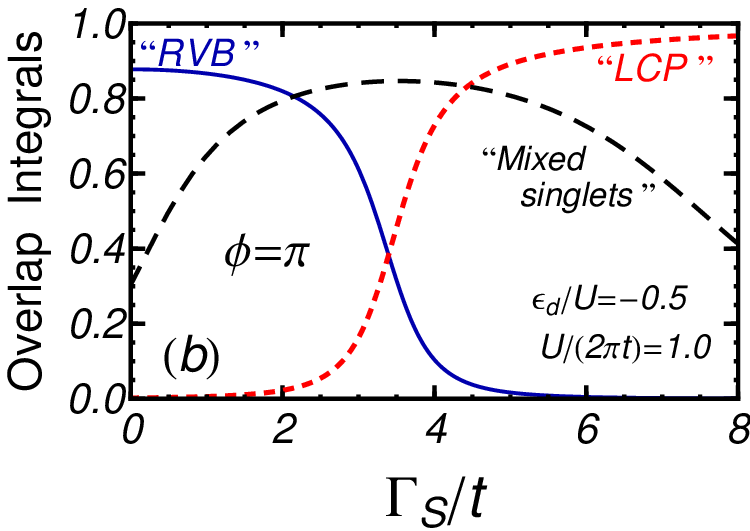}
\end{minipage}
\caption{
 Overlap integrals,       
 $\bigl|\langle \Phi_{SB} | \Psi_\mathrm{1st}^{S=1/2}  \rangle\bigr|^2$,
 $\big|\langle \Phi_{RVB} | \Psi_\mathrm{1st}^{S=1/2} \rangle\bigr|^2$,
 $\big|\langle \Phi_{LCP} | \Psi_\mathrm{1st}^{S=1/2} \rangle\bigr|^2$, 
 and 
 $\sum_{j=1}^2 
 \left|\langle \Phi_{\mathrm{mix},j} | \Psi_\mathrm{1st}^{S=0} 
 \rangle\right|^2$,  
are plotted  
 vs $\Gamma_S$ for  (a) $\phi = 0$ and  (b) $\phi = \pi$. 
$|\Psi_\mathrm{1st}^{S=0} \rangle$ and  
$|\Psi_\mathrm{1st}^{S=1/2} \rangle$ are   
the lowest-energy states 
in the $S=0$ and $1/2$ subspaces, respectively. 
} 
\label{fig:overlap_integral}
\end{figure}

\begin{figure}[t] 
 \leavevmode

\rule{0.025\linewidth}{0cm}
\begin{minipage}{0.9\linewidth}
\raisebox{0.2cm}{
\includegraphics[width=0.48\linewidth]{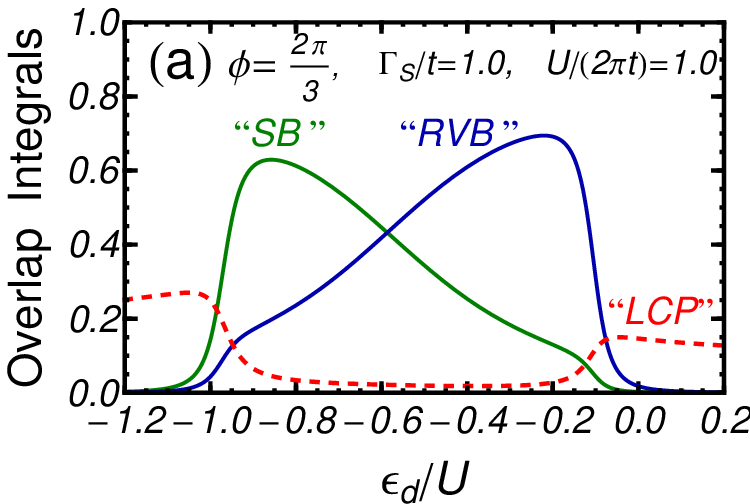}
}
\rule{0.02\linewidth}{0cm}
\includegraphics[width=0.47\linewidth]{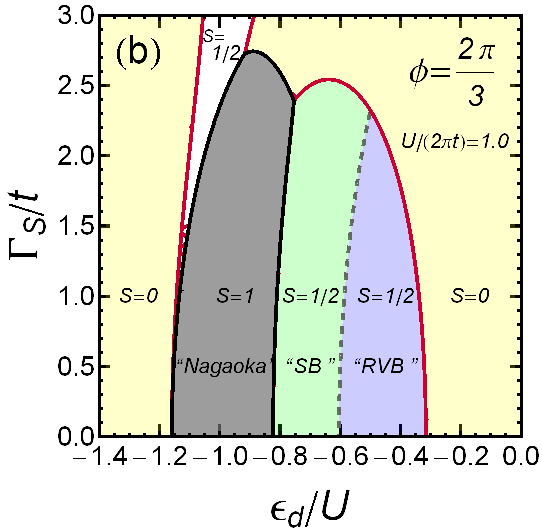}
\end{minipage}
\caption{
Grand states  for $\phi = 2\pi/3$ and $U=2\pi t$. 
(a): $\epsilon_d$ dependence of  overlap integrals at $\Gamma_S/t=1.0$. 
(b): Enlarged phase diagram, 
in which the dashed line traces the points where  
the overlap integral for the SB and that for RVB coincide, as   
$\bigl|\langle \Phi_{SB} | \Psi_\mathrm{1st}^{S=1/2}  \rangle\bigr|^2=
\big|\langle \Phi_{RVB} | \Psi_\mathrm{1st}^{S=1/2} \rangle\bigr|^2$.
} 
\label{fig:overlap_integral_phi066}
\end{figure}


Infinitesimal SC proximity lifts this degeneracy of the 
singlet bond (SB) and the resonating valence bond (RVB) states 
as seen in Fig.\ \ref{fig:energy} which shows $\Gamma_S$ dependence 
of the eigenvalues near half-filling $\epsilon_d=-0.5U$ 
for $U=2\pi t$.  
Detailed features of the eigenvectors can be deduced from  
the overlap integrals with the fixed-point wavefunctions defined in 
 Eqs.\ \eqref{eq:SB_fixed_point}--\eqref{eq:Mixed_singlet_fixed_point}: 
\begin{align}
 \bigl|\langle \Phi_{SB} | \Psi_\mathrm{1st}^{S=1/2}  \rangle\bigr|^2,
\quad \ 
 \big|\langle \Phi_{RVB} | \Psi_\mathrm{1st}^{S=1/2} \rangle\bigr|^2,
\quad \ 
 \big|\langle \Phi_{LCP} | \Psi_\mathrm{1st}^{S=1/2} \rangle\bigr|^2, 
\quad \ 
 \sum_{j=1}^2 
 \left|\langle \Phi_{\mathrm{mix},j} | \Psi_\mathrm{1st}^{S=0} 
 \rangle\right|^2 .
\end{align}
Here, $|\Psi_\mathrm{1st}^{S=0} \rangle$ and  
$|\Psi_\mathrm{1st}^{S=1/2} \rangle$ are   
the lowest eigenstates 
in the $S=0$ and $1/2$ subspaces, respectively. 
Some typical examples are shown in Fig. \ref{fig:overlap_integral}. 
For small proximities $\Gamma_S/t \lesssim 2.0$, 
the singlet-bond state $\left| \Phi_{SB} \right\rangle$ 
 dominates the ground state for $\phi=0$, 
whereas the RVB dominates for $\phi =\pi$. 
Therefore, in this region the ground state must evolve   
 from the SB to the RVB as $\phi$ increases, 
as it  will be discussed later.   

The overlap integrals 
 $\bigl|\langle \Phi_{SB} | \Psi_\mathrm{1st}^{S=1/2}  \rangle\bigr|^2$ 
and 
 $\bigl|\langle \Phi_{RVB} | \Psi_\mathrm{1st}^{S=1/2}  \rangle\bigr|^2$ 
  decrease as  $\Gamma_S/t$ increases, and 
a crossover occurs for both $\phi=0$ and $\pi$.
Then, for large proximities $\Gamma_S/t \gtrsim 4.0$, 
the local Cooper pairing $\left| \Phi_{LCP} \right\rangle$, 
in which the SC proximities  
at the bottom and the local spin at the apex of 
the triangle coexist, determines the ground-state properties.  
We can also see in the intermediate region near $\Gamma_S/t \simeq 3.0$ 
that the lowest singlet eigenstate $|\Psi_\mathrm{1st}^{S=0} \rangle$ 
has a large overlap with 
the mixed singlet states 
 $\left| \Phi_{\mathrm{mix},j} \right\rangle$,  
in which the local Cooper pair at the bottom and 
a singlet bond in the oblique side share the dots 
in the triangular configuration. 
This type of the singlet becomes the ground state 
in the center of the phase diagrams in 
Figs.\ \ref{fig:results_2SC_terminal} (a)--(d), 
and it explains why the $S=0$ region expands as $\phi$ increases 
for intermediate $\Gamma_S$ near half-filling. 
This is also consistent with the behavior of the eigenvalues 
seen in Fig.\ \ref{fig:energy} (b), in which     
$|\Psi_\mathrm{1st}^{S=0} \rangle$  becomes the ground state 
around  $\Gamma_S/t \simeq 3.0$ for $\phi=\pi$.

For small $\Gamma_S$ near half-filling, 
the ground state must change from the SB to the RVB 
at finite $\phi$, as mentioned.  
We see in Fig.\ \ref{fig:overlap_integral_phi066} (a) 
that this occurs near $\phi \simeq 2\pi/3$ for $U=2\pi t$. 
As $\epsilon_d$ increases, the fixed-point state showing the greatest 
overlap changes from the SB to the RVB 
near $\epsilon_d/U \simeq -0.6$.    
Therefore, the magnetic $S=1/2$ ground state 
can be classified according to these overlap integrals. 
Figure \ref{fig:overlap_integral_phi066} (b) shows  
an enlarged phase diagram for $\phi = 2\pi/3$. 
The dashed line represents the boundary, on which    
$\bigl|\langle \Phi_{SB} | \Psi_\mathrm{1st}^{S=1/2}  \rangle\bigr|^2=
\big|\langle \Phi_{RVB} | \Psi_\mathrm{1st}^{S=1/2} \rangle\bigr|^2$. 
This boundary line penetrates into the $S=1/2$ region  
from the adjacent $S=0$ region on the right,   
and will move towards the $S=1$ Nagaoka region on the left 
as $\phi$ increases further.

\section{Bogoliubov zero-energy excitation} 

\begin{figure}[t]
\leavevmode

\rule{0.05\linewidth}{0cm}
\begin{minipage}{0.9\linewidth}
\leavevmode
\includegraphics[width=0.47\linewidth]{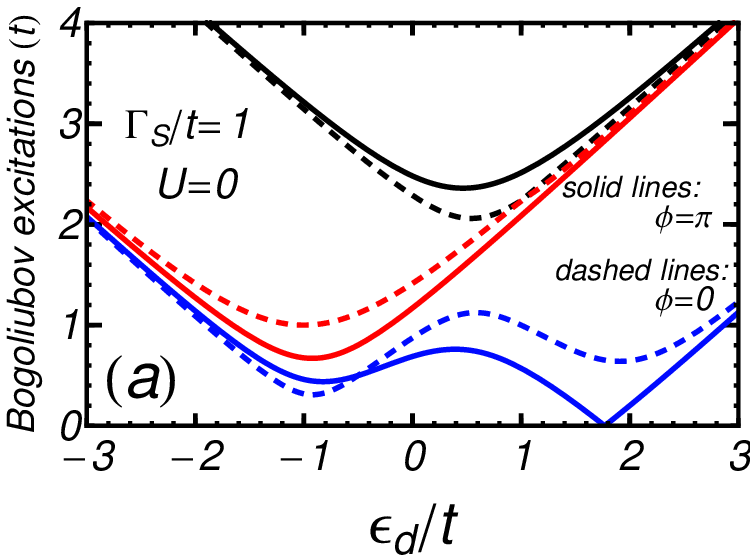}
\rule{0.02\linewidth}{0cm}
\includegraphics[width=0.49\linewidth]{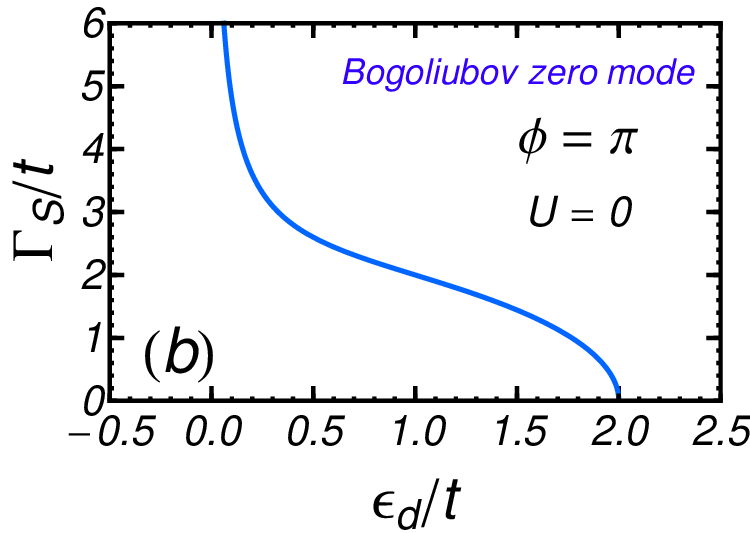}
\end{minipage}
\caption{
 (a): Bogoliubov excitations $E_{k}^{}$ vs $\epsilon_d$ for $U=0$ 
and $\Gamma_S/t=1.0$: the solid and dashed lines represent 
the energy levels for $\phi=\pi$ and $\phi=0$, respectively.
(b): Zero-energy mode of $E_{k}^{}=0$ emerges at $\phi=\pi$   
along the blue line in the $\epsilon_d$ vs $\Gamma_s$ plane. 
}
\label{fig:phase_diagram_bogoliubov_zero_mode}
\end{figure}

Another characteristics of the ground-state phase diagrams,  
shown in Fig.\ \ref{fig:results_2SC_terminal} for $U=2\pi t$,  
 is that the boundary between the $S=0$ and $S=1/2$ regions 
at $\epsilon_d/U \simeq 0.3$ is connected 
for large $\phi$ ($\gtrsim 0.56 \pi$),   
while the two regions are separated 
for small $\phi$ ($\lesssim 0.55 \pi$).  
Note that this change takes place for small electron fillings: 
the quantum dots are almost empty 
on the right side of this phase boundary, 
and the first electron enters into the TTQD 
as $\epsilon_d$ passes through the boundary.
Therefore, around this boundary   
the Coulomb interaction $U$ can be treated perturbatively,    
and thus the Bogoliubov quasiparticles $\gamma_{k\sigma}^{}$ 
which diagonalize the noninteracting Hamiltonian  
play an essential role, 
\begin{align}
H_\mathrm{eff}^{} \, \xrightarrow{U= 0\ } \, \sum_{k=1}^3 \sum_{\sigma} 
E_{k}^{}\,
\gamma_{k\sigma}^{\dag} \gamma_{k\sigma}^{}
+ \mathrm{const.} 
\end{align}
Figure \ref{fig:phase_diagram_bogoliubov_zero_mode} (a) 
shows the three energy levels of $E_{k}^{}$ for 
 $\phi=0$ (dashed line) and  $\phi=\pi$ (solid line).
We see that the lowest excitation energy becomes zero 
for $\phi = \pi$ at $\epsilon_d/t \simeq 1.88$ and $\Gamma_S = t$.  
The zero mode, where $E_{k}^{}=0$, 
 emerges in the parameter space for $\phi=\pi$ along the line,
\begin{align}
&
\Gamma_S^2 \, \epsilon_d   
+ (\epsilon_d-2t) (\epsilon_d+t)^2 = 0 
\;.
\end{align}
On this line, 
which is also shown in Fig.\ \ref{fig:phase_diagram_bogoliubov_zero_mode} (b), 
the excitation has a four-fold degeneracy that can be classified 
according to the occupation number of the zero mode. 
The Coulomb interaction $U$ lifts this degeneracy, 
and then a singly occupied $S=1/2$ doublet becomes  ground state 
in the region around the line of the zero-energy excitations.
This explains the characteristics of the phase diagrams, 
Fig.\ \ref{fig:results_2SC_terminal} (c) and (d), 
near $\phi=\pi$: 
the boundary between the $S=0$ and $S=1/2$ regions 
near $\epsilon_d/U \simeq 0.3$ does not enclose itself but  
stretches out towards the direction of positive large $\Gamma_S$'s.

\section{Summary}
We have studied in detail the ground states 
of the TTQD embedded in a Josephson junction 
between two SC leads. Specifically, 
various quantum phases emerging in a wide parameter space  
have been classified according to the fixed points 
which include new ones, the LCP and the mixed singlets, 
describing the SC proximities. 
The results of the overlap integrals clarify  
the correspondence between the ground state and 
the fixed points, 
and give a clear physical interpretation 
to the phase diagram. 
Effects of the normal lead on low-energy properties 
will be discussed elsewhere.

\ack
We would like to thank 
Yasuhiro Yamada and Rui Sakano for discussions.
This work is supported by the JSPS Grant-in-Aid for 
 Scientific Research (C) (No.\ 26400319) 
and (S) (No.\ 26220711). 
Numerical computation was partly carried out 
at Yukawa Institute Computer Facility.


\section*{References}
\begin{thebibliography}{9}


\bibitem{KKA1}
T.~Kuzmenko, K.~Kikoin and Y.~Avishai, 
Phys.\ Rev.\ Lett.\  {\bf 96},  046601 (2006).


\bibitem{ONTN}
A.~Oguri,   Y.~Nisikawa, Y.~Tanaka, and T.~Numata, 
J.\ Magn.\ \& Magn.\ Mater.\ {\bf 310}, 1139 (2007).




\bibitem{Zitko2}
R.~\v{Z}itko and J.~Bon\v{c}a, 
Phys. Rev. B {\bf 77}, 245112 (2008). 



\bibitem{Logan3}
A.~K.~Mitchell, T.~F.~Jarrold, and D.~E.~Logan, 
Phys. Rev. B {\bf 79}, 085124 (2009). 



\bibitem{Ulloa}
E.~Vernek, C.~A.~B\"{u}sser, 
G.~B.~Martins, E.~V.~Anda, N.~Sandler and S.~E.~Ulloa, 
Phys. Rev. B {\bf 80}, 035119 (2009).

\bibitem{Numata2}
T.~Numata, Y.~Nisikawa, A.~Oguri, and A.~C.~Hewson,   
Phys. Rev. B {\bf 80}, 155330 (2009). 


\bibitem{aoTTQD2011}
  A.\ Oguri, S.\ Amaha, Y.\ Nishikawa, 
  T.\ Numata, M.\ Shimamoto, A.\ C.\ Hewson and  S.\ Tarucha, 
  Phys.\ Rev.\ B {\bf 81}, 075404 (2011).

\bibitem{Koga}
 M.\ Koga, M.\ Matsumoto, and H.\ Kusunose, 
 J.\ Phys.\ Soc.\ Jpn.\ {\bf 82}, 093706 (2013).





\bibitem{MartinYeyati}
A.\ Mart$\acute{\i}$n-Rodero and A.\ Levy Yeyati, 
Adv.\ Phys.\ {\bf 60}, 899 (2011).


\bibitem{sc3} 
J.\ Bauer, A.\ Oguri, and  A.\ C.\ Hewson,  
J.\ Phys.:\ Condes.\ Mat.\  {\bf 19}, 486211 (2007).

\bibitem{TanakaNQDS}
Yoichi Tanaka, N.\ Kawakami, and A.\ Oguri,
J.\ Phys.\ Soc.\ Jpn.\ \textbf{76}, 074701 (2007);
\textbf{77}, 098001(E) (2008).


\bibitem{Deacon}
R.\ S.\ Deacon, Yoichi Tanaka, A.\ Oiwa, R.\ Sakano, K.\ Yoshida, K.\ Shibata,
 K.\ Hirakawa, and S.\ Tarucha, 
Phys.\ Rev.\  B {\bf 81}, 121308 (2010).



\bibitem{MaurandSchonenberger}
R.\ Maurand and C.\ Sch\"{o}nenberger,
Physics {\bf 6}, 75 (2013).


\bibitem{GovePalaKonig_PRB}
M.\ Governale, M.\ G.\ Pala,  and J.\ K\"{o}nig,   
Phys.~Rev.~B {\bf 77}, 134513 (2008).  


\bibitem{aoYoichiBauer}
 A.\ Oguri, Yoichi Tanaka, and J.\ Bauer, 
Phys.~Rev.~B {\bf 87}, 075432 (2013).




\end{thebibliography}
\end{document}